\documentclass[superscriptaddress,twocolumn,english,floatfix,aps,prd,amsmath,amssymb,longbibliography]{revtex4-2}
\usepackage[T1]{fontenc}
\usepackage[utf8]{inputenc}
\usepackage{babel}
\usepackage{comment}
\usepackage{color}
\usepackage{times}
\usepackage{epsfig}
\usepackage{subfigure}
\usepackage{amsmath}
\usepackage{amssymb}
\usepackage{graphicx}
\usepackage{float}
\usepackage[colorlinks=true]{hyperref}
\hypersetup{citecolor=blue,linkcolor=blue,urlcolor=blue}

\begin{document}

\title{Renormalization of the band gap in 2D materials near an interface between two dielectrics}

\author{Alessandra N. Braga}
\email{alessandrabg@ufpa.br}
\affiliation{Campus de Ananindeua, Universidade Federal do Par\'{a}, 67130-660, Ananindeua,
Brazil}
\author{Wagner P. Pires}
\email{wagner.pires@ufopa.edu.br }
\affiliation{Instituto de Ci\^{e}ncias da Educa\c{c}\~{a}o, Universidade Federal do Oeste do
Par\'{a}, 68040-255, Santar\'{e}m, Brazil}
\author{Jeferson Danilo L. Silva}
\email{jdanilo@ufpa.br}
\affiliation{Campus Salin\'{o}polis, Universidade Federal do Par\'{a}, 68721-000, Salin\'{o}polis,
Brazil}
\author{Danilo T. Alves}
\email{danilo@ufpa.br}
\affiliation{Faculdade de F\'{i}sica, Universidade Federal do Par\'{a}, 66075-110, Bel\'{e}m,
Brazil}
\author{Van S\'{e}rgio Alves}
\email{vansergi@ufpa.br}
\affiliation{Faculdade de F\'{i}sica, Universidade Federal do Par\'{a}, 66075-110, Bel\'{e}m, Brazil}
\date{\today}
%
\begin{abstract}
We investigate how the renormalization of the band gap in a planar 2D material is affected by the consideration of two nondispersive semi-infinite dielectrics, with dielectric constants $\epsilon_1$ and $\epsilon_2$, separated by a planar interface.
Using the pseudo quantum electrodynamics to model the Coulomb interaction between electrons, we show how the renormalization of the band gap depends on $\epsilon_1$ and $\epsilon_2$, and also of the distance between the 2D material and the interface between the two dielectrics.
In the appropriate limits, our results reproduce those found in the literature for the band gap renormalization when a single dielectric medium is considered.
\end{abstract}

\maketitle

\section{Introduction}

Quantum field theory in $2+1$ dimensions has been employed to describe various aspects of high-energy physics, including quark confinement and chiral symmetry breaking \cite{Maris1994,*Appelquist1985,*Burden1992,*Grignani1996,*Maris1995}, as well as condensed matter physics phenomena like the quantum Hall effect and superconductivity \cite{Range-1990,*Wilczek1990,*Halperin1982,*Laughlin1981,*Laughlin1988,*Chen1989}. Since the discovery of graphene \cite{Geim2007}, a two-dimensional material with the thickness of a carbon atom and a massless relativistic-like dispersion relation, numerous attempts have been made to develop theoretical models capable of accurately describing its remarkable properties. Other materials with a honeycomb-like lattice, similar to graphene, have gained attention due to their potential technological applications \cite{Radisavljevic2011,*Wang2012,*Liu2014}. Silicene, phosphorene and transitional metal dichalcogenides (TMDs) are examples of such materials that primarily differ from graphene in that their low-energy excitations are, approximately, described by the massive Dirac equation \cite{Castro-RMP-2009}. 

It is well known that electromagnetic interaction plays an important role in the transport properties of these materials. Studies on graphene reveal that  measurement of the renormalization of the Fermi velocity \cite{Geim-2011-Nature}, the direct measurement of the dc conductivity \cite{Du2008},  and the experimental observation of the fractional quantum Hall effect in ultraclean samples \cite{Du2009,*Bolotin2009,*Ghahari2021,*Dean2011} clearly demonstrate that electromagnetic interaction is in fact important, at least for a certain temperature scale.

A theoretical model that accurately describes the electronic properties of these materials must account for the fact that electrons and photons exist in different spacetime dimensions. In these materials, electrons exist in $2+1$ dimensions while photons exist in $3+1$ dimensions. Therefore, constructing a quantum field theory model requires incorporating mixed dimensions. In 1993, Marino proposed a model with these characteristics, which represents QED4 projected onto  a two-dimensional  plane. This theory is known as pseudo quantum electrodynamics (PQED) \cite{Marino-NPB-1993} and it is sometimes referred to as reduced quantum electrodynamics \cite{Gorbar-PRD-2001}. 

The PQED model has been demonstrated to exhibit unitarity \cite{Marino-PRD-2014}, causality \cite{Marino-1992}, as well as being scale invariant for a massless theory \cite{Dudal2019,Heydeman2020}. In addition, it reproduces the static Coulombian potential, instead of the peculiar logarithmic one from QED in $2+1$ dimensions. As a result, it has been widely employed with significant success in numerous scenarios involving the aforementioned two-dimensional materials \cite{Menezes2016,VanSergio-PRB-2017,Marino2018,Pedrelli-2020}. 

\begin{figure}[h]
	\centering
	\epsfig{file=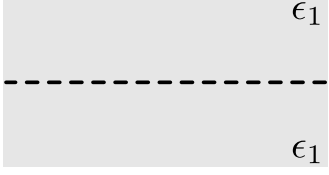,  width=0.5\linewidth}
	\caption{Illustration of a 2D material (dashed line) embedded in a dielectric with dielectric constant $\epsilon_1$.} 
	\label{fig:luis}
\end{figure}

Recently, a branch of PQED that includes effects of boundary conditions imposed by interfaces to the electromagnetic field, called cavity PQED, has established that the renormalization of the Fermi velocity and the transport properties of graphene are significantly altered by the presence of a grounded metal plate or a cavity in close proximity to a graphene sheet \cite{Pires-NPB-2017,Pires-NPB-2018,Pedrelli-2020}.
The PQED has also been employed in the random-phase approximation (RPA) to calculate the renormalized mass $m$ (energy gap) as a function of the carrier concentration $n$ in TMD materials, embedded in a dielectric medium with dielectric constant $\epsilon_1$ \cite{Fernandez-etal-PRD-2020} (see Fig. \ref{fig:luis}). These authors found that, given $m_0\equiv m(n_0)$ as a reference value provided by experiments, the curve $m(n)$ is given by
\begin{equation}
m=m_{0}\left(\frac{n}{n_{0}}\right)^{C_{\lambda}/2},
\label{eq:Luis}
\end{equation}
where
\begin{equation}
C_{\lambda}=-\frac{1}{\pi^{2}}\left[4+\frac{4\cos^{-1}\left(\lambda\right)}{\lambda\sqrt{1-\lambda^{2}}}-\frac{2\pi}{\lambda}\right],
\label{eq:Luis-C}
\end{equation}
with $\lambda\equiv e^{2}/(16\epsilon_{1}v_{F})$, $e$ the electrical charge, and $v_{F}$ the Fermi velocity. 
Theoretical results obtained in Ref. \cite{Fernandez-etal-PRD-2020}, by applying Eq. \eqref{eq:Luis} to tungsten diselenide (WSe$_2$) \cite{Nguyen2019} and molybdenum disulfide (MoS$_2$), are in excellent agreement with experimental data \cite{Nguyen2019,Liu2019}, reinforcing the usefulness of PQED in the study of the electronic properties of these materials. 
\begin{figure}[h]
	\centering
	\epsfig{file=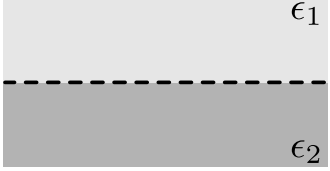,  width=0.5\linewidth}
	\caption{Illustration of a 2D material (dashed line) in the interface between two dielectrics with dielectric constants $\epsilon_1$ and $\epsilon_2$.} 
	\label{fig:luis-ext}
\end{figure}

\begin{figure}[h]
	\centering
	\epsfig{file=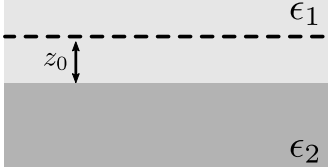,  width=0.5\linewidth}
	\caption{Illustration of a 2D material (dashed line) near to the flat interface between two dielectrics with dielectric constants $\epsilon_1$ and $\epsilon_2$. The distance between the material and the interface is $z_0$, and the 2D material is immersed in the medium with dielectric constant $\epsilon_1$.} 
	\label{fig:nos}
\end{figure}

Although Eq. \eqref{eq:Luis} can be applied even when the 2D material is in the interface between two dielectrics with dielectric constants $\epsilon_1$ and $\epsilon_2$ [making $\epsilon_1\to (\epsilon_1+\epsilon_2)/2$ in Eq. \eqref{eq:Luis}], as illustrated in Fig. \ref{fig:luis-ext}, this
formula is unable to address the situation where the 2D material is at a distance $z_0$ from the interface (as illustrated in Fig. \ref{fig:nos}).
In the present paper, we investigate, in the context of the cavity PQED, the effect of a planar interface, separating two nondispersive semi-infinite dielectrics  ($\epsilon_1$ and $\epsilon_2$), on the renormalization of the mass (band gap) in a planar 2D material located at a distance $z_0$ from the interface (Fig. \ref{fig:nos}). 
When $z_0=0$ (Fig. \ref{fig:luis-ext}) or $\epsilon_1=\epsilon_2$ (Fig. \ref{fig:luis}), the
formula to be obtained here recovers Eq. \eqref{eq:Luis} found in Ref. \cite{Fernandez-etal-PRD-2020}.

This paper is organized as follows. In Sec. \ref{sec-model}, we present the model and its Feynman rules. 
In Sec. \ref{sec-photon-propagator}, we calculate the photon propagator considering the influence of the dielectrics and, subsequently, we incorporate the effects of the polarization tensor into this propagator. In Sec. \ref{sec-electron-self-energy}, we calculate the electron self-energy at one-loop order in RPA. In Section \ref{sec-renormalization-group-functios}, we obtain the renormalization group functions based on the findings from the preceding section, and we derive the renormalization of the band gap within the framework of the RPA. In Sec. \ref{sec-application}, we apply our result to investigate the renormalization of the band gap for a monolayer of WSe$_2$, when this 2D system is situated within a boron nitride substrate, with a distance $z_0$ separating it from the vacuum interface, or conversely, when it is located within the vacuum with a distance $z_0$ from the interface of a boron nitride substrate. Finally, in Sec. \ref{sec-final-remarks} we present our final remarks. We also include an Appendix with some details of the calculations regarding the gauge field propagator in the static regime when dielectrics are present.

\section{The model and the Feynman Rules}
\label{sec-model}
The effective theory and complete description in $2 + 1$ dimensions for electronic systems moving on a plane, but interacting as particles in $3 + 1$ dimensions is given by \cite{Marino-NPB-1993}
\begin{equation}
	\mathcal{L}_{\mathrm{PQED}} = \frac{1}{2}\frac{{F}_{\mu\nu}{F}^{\mu\nu}}{(-\square)^{1/2}}
	+\mathcal{L}_{\text{D}} + j^{\mu}A_{\mu}
	-\frac{\xi}{2}{A}_\mu\frac{\partial^\mu\partial^\nu}{(-\square)^{1/2}}A_\nu,
	\label{L-PQED}
\end{equation}
where $\square$ is the d'Alembertian operator, $F^{\mu\nu}$ is the usual field intensity tensor of the gauge field $A_{\mu}$ and we consider $c=\hbar=1$. 
To emulate 2D materials, we shall consider an anisotropic version of the Dirac Lagrangian given by $\mathcal{L}_{\mathrm{D}} = \bar{\psi}_a\left(i\gamma^0\partial_0+iv_F\boldsymbol{\gamma}\cdot\nabla\right)\psi_a$, where $\bar{\psi}_a=\psi^{\dagger}_{a}\gamma^0$, and $a=1,\dots,N$ is a flavor index representing a sum over valleys $K$ and $K^{\prime}$, $\gamma^{\mu}$ are rank-4 Dirac matrices and  $\psi^\dagger_a = \big(\psi^{\star}_{A\uparrow}\;\psi^{\star}_{A\downarrow}\;\psi^{\star}_{B\uparrow}\; \psi^{\star}_{B\downarrow}\big)_a$ is a four-component Dirac spinor representing electrons in sublattices $A$ and $B$ in two-dimensional system, with different spin orientations. The last term corresponds to the gauge fixing term.

From Eq. (\ref{L-PQED}), one obtains the free photon propagator in Euclidean space,
\begin{equation}
	\Delta_{\mu\nu}^{(0)}(k)=\frac{1}{2\sqrt{k^2}}\left[\delta_{\mu\nu}-\left( 1-\frac{1}{\xi}\right) \frac{k_{\mu}k_{\nu}}{k^2}\right],
	\label{eq:gauge-prop-completo}
\end{equation}
where $k_{\mu}=(k_0,\mathbf{k})$ and $\mathbf{k}=(k_1,k_2)$. In the nonretarded regime, considering the Feynman gauge ($\xi=1$), it becomes
\begin{equation}
	\Delta_{\mu\nu}^{(0)}(k_0=0,|\mathbf{k}|)=\frac{1}{2|\mathbf{k}|}\delta_{0\mu}\delta_{0\nu},
	\label{eq:gauge-prop-free}
\end{equation}
which leads to the Coulombian potential for static charges (instead of the peculiar logarithmic one from QED in $2+1$ dimensions),
\begin{equation}
	V(|\mathbf{r}|)=\frac{e}{4\pi}\frac{1}{|\mathbf{r}|}.
	\label{eq:usual-coulombian}
\end{equation}

The fermion propagator is
\begin{equation}
	S_{F}^{(0)}(k^{\mu})=\frac{k_{0}\gamma_{0}+v_{F}\mathbf{k}\cdot\boldsymbol{\gamma}+m}{k_{0}^{2}+v_{F}^{2}|\mathbf{k}|^{2}+m^{2}},
\end{equation}
and the vertex interactions is $\gamma^{\mu}=(\gamma^0, v_F\gamma^{i})$.

\begin{figure}[h]
	\centering
	\epsfig{file=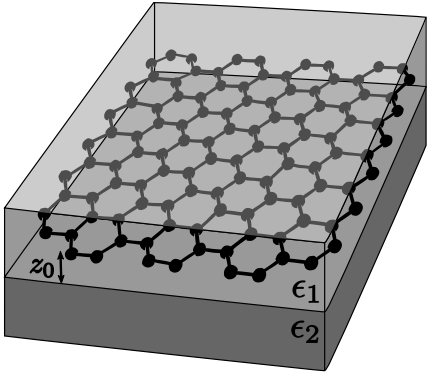,  width=0.65 \linewidth}
	\caption{Illustration of a 2D material parallel and near to the interface between two dielectrics with dielectric constants $\epsilon_1$ and $\epsilon_2$. The distance between the material and the interface is $z_0$, the 2D material is immersed in the medium with dielectric constant $\epsilon_1$.}
	\label{modelo}
\end{figure}

In this paper, we consider a planar 2D material in a dielectric medium with dielectric constant $\epsilon_1$. The 2D material is separated by a distance $z_0$ from a parallel interface with another dielectric medium with dielectric constant $\epsilon_2$ (see Figs. \ref{fig:nos} and \ref{modelo}).

\section{The photon propagator}
\label{sec-photon-propagator}

The gauge-field propagator in the static regime, considering the two dielectric media, is given by (see the Appendix)
\begin{equation}
\Delta_{00}^{(0)}(\left|\mathbf{k}\right|)=\frac{1}{2\epsilon_{1}\sqrt{|\mathbf{k}|^{2}}}\left[1-\frac{\epsilon_{2}-\epsilon_{1}}{\epsilon_{2}+\epsilon_{1}}\exp{\left(-2z_{0}|\mathbf{k}|\right)}\right].
\label{eq:gauge-prop-dieletrico}
\end{equation}
In the following, we shall use this equation to obtain the new expression for the propagator in the
large $N$ expansion.

\subsection{Photon propagator in large $N$ expansion}

We consider the large $N$ expansion at one-loop order approximation, which is equivalent to the RPA \cite{Vozmediano-2012,Vozmediano-PRB-2010}. This approximation has been used in the description of some properties of suspended \cite{Vozmediano-PRB-1999,CastroNeto-PRB-2009} and doped \cite{Das-Sarma-PRB-2007,Polini-science-2007,Hwang-PRB-2007} graphene. It can be conveniently implemented by replacing $e\rightarrow e/\sqrt{N}$, for a fixed $e$.

\begin{figure}[b]
	\centering
	\epsfig{file=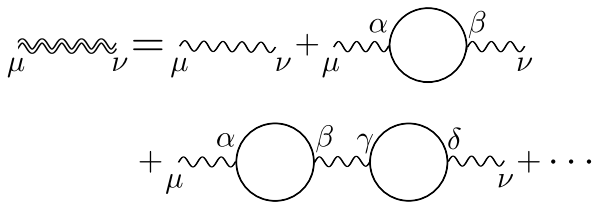,  width=0.9 \linewidth}
	\caption{Gauge field propagator in dominant order in the $1/N$ expansion.}
	\label{feymann_diagram}
\end{figure}

We will consider the geometric series to calculate the full propagator of the gauge-field. Assuming that the interaction vertex is just given by $\gamma_{0}$ (static regime), as illustrated in the Fig. \ref{feymann_diagram},
we can write
\begin{equation}
\Delta_{00}^{\mathrm{RPA}}(k)=\Delta_{00}^{(0)}\left(1-\Pi^{00}\Delta_{00}^{(0)}\right),
\end{equation}
where \cite{Fernandez-etal-PRD-2020}
\begin{equation}
\Pi^{00}=-\frac{e^{2}}{8}\left[\frac{|\mathbf{k}|^{2}}{\sqrt{k_{0}^{2}+v_{F}^{2}\mathbf{k}^{2}}}-\frac{4|\mathbf{k}|^{2}m^{2}}{(k_{0}^{2}+v_{F}^{2}|\mathbf{k}|^{2})^{3/2}}\right].
\end{equation}
Therefore, we find
\begin{eqnarray}
	\Delta_{00}^{\mathrm{RPA}}(k) & = & \frac{1}{2\epsilon_{1}|\mathbf{k}|}\left\{ \frac{1}{1-\kappa_{21}\exp(-2z_{0}|\mathbf{k}|)}+\frac{1}{16}\frac{e^{2}}{\epsilon_{1}}\right.\nonumber \\
	&  & \left.\times\left[\frac{|\mathbf{k}|}{\sqrt{k_{0}^{2}+v_{F}^{2}\mathbf{k}^{2}}}-\frac{4|\mathbf{k}|m^{2}}{(k_{0}^{2}+v_{F}^{2}|\mathbf{k}|^{2})^{3/2}}\right]\right\} ^{-1}.\nonumber \\
	\label{propagador-RPA}
\end{eqnarray}
where we defined
\begin{equation}
\kappa_{21}=\frac{\epsilon_{2}-\epsilon_{1}}{\epsilon_{2}+\epsilon_{1}}.
\end{equation}
Note that, in the particular case where $\epsilon_{1}=\epsilon_{2}$ in Eq. (\ref{propagador-RPA}) our model recovers the case of the full propagator of the gauge-field within a
single medium \citep{Fernandez-etal-PRD-2020}.

\section{Electron self-energy}
\label{sec-electron-self-energy}

\begin{figure}[t]
	\centering
	\epsfig{file=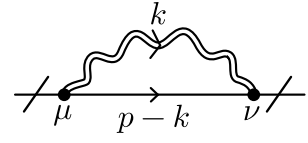,  width=0.45 \linewidth}
	\caption{Electron self-energy diagram.}
	\label{self}
\end{figure}

Considering the static approximation, the electron self-energy, represented in Fig. \ref{self}, reads
\begin{equation}
	\Sigma(p)=\frac{e^{2}}{N}\int\frac{\mathrm{d}^{3}k}{(2\pi)^{3}}\gamma_{0}S_{F}^{(0)}(p_{\mu}-k_{\mu})\gamma_{0}\Delta_{00}^{\mathrm{RPA}}(k).
\end{equation}
Assuming that the Dirac matrices satisfy $\{\gamma^{\mu},\gamma^{\nu}\}=-2\delta^{\mu\nu}$, $\gamma^{0}\gamma^{0}=-\mathbb{I}$ and $\gamma^{0}\boldsymbol{\gamma}\gamma^{0}=\boldsymbol{\gamma}$,  
and performing a power series expansion in the external momentum, we obtain
\begin{equation}
	\Sigma(p)=\mathcal{Z}_{m}+\mathcal{Z}_{0}\gamma_{0}p_{0}+v_{F}\mathcal{Z}_{1}\gamma_{i}p_{i},\label{eq:auto-energia}
\end{equation}
where the dominant terms are
\begin{equation}
	\mathcal{Z}_{m}=-\frac{e^{2}}{N}\int\frac{\mathrm{d}^{3}k}{(2\pi)^{3}}\frac{m}{k^{2}+m^{2}}\Delta_{00}^{\mathrm{RPA}}(k),
	\label{eq:zm-RPA}
\end{equation}
\begin{equation}
	\mathcal{Z}_{0}=-\frac{e^{2}}{N}\int\frac{\mathrm{d}^{3}k}{(2\pi)^{3}}\frac{v_{F}^{2}\left|\mathbf{k}\right|^{2}-k_{0}^{2}+m^{2}}{(k^{2}+m^{2}){}^{2}}\Delta_{00}^{\mathrm{RPA}}(k),
	\label{eq:z0-RPA}
\end{equation}
and
\begin{equation}
	\mathcal{Z}_{1}=\frac{e^{2}}{N}\int\frac{\mathrm{d}^{3}k}{(2\pi)^{3}}\frac{k_{0}^{2}+m^{2}}{\left(k^{2}+m^{2}\right)^{2}}\Delta_{00}^{\mathrm{RPA}}(k).
	\label{eq:z1-RPA}
\end{equation}
Then, we perform a variable change $v_{F}k_{i}\rightarrow\bar{k}_{i}$
in Eq. (\ref{eq:auto-energia}) to spherical coordinates, the lowest-order terms can be written as
\begin{equation}
	\mathcal{Z}_{m} = -\frac{2\bar{\lambda}}{\pi{}^{2}N}\int_{0}^{\pi}\mathrm{d}\theta\int_{0}^{\Lambda v_F}k\,\mathrm{d}k\frac{m}{k^{2}+m^{2}}f(k,\theta),
	\label{eq:zm}
\end{equation}
\begin{equation}
	\mathcal{Z}_{0}=\frac{2\bar{\lambda}}{\pi{}^{2}N}\int_{0}^{\pi}\mathrm{d}\theta\int_{0}^{\Lambda v_F}k\,\mathrm{d}k\frac{k^{2}(1-2\sin^{2}\theta)-m^{2}}{(k^{2}+m^{2})^{2}}f(k,\theta),
	\label{eq:z0}
\end{equation}
\begin{equation}
	\mathcal{Z}_{1}  =  \frac{2\bar{\lambda}}{\pi{}^{2}N}\int_{0}^{\pi}\mathrm{d}\theta\int_{0}^{\Lambda v_F}k\,\mathrm{d}k\frac{k^{2}\cos^{2}\theta+m^{2}}{(k^{2}+m^{2})^{2}}f(k,\theta),
	\label{eq:z1}
\end{equation}
where $\bar{\lambda}=e^{2}/16\epsilon_{1}v_{F}$ and we introduced an ultraviolet cut-off $\Lambda v_F$ in the integrals above.
We also defined
\begin{eqnarray}
f(k,\theta) & = & \left\{ [1-\kappa_{21}\exp(-2z_{0}k\sin\theta/v_{F})]^{-1}\right.\nonumber \\
 &  & \left.+\bar{\lambda}\left(1-4m^{2}/k^{2}\right)\sin\theta\right\} ^{-1}.
\end{eqnarray}

\section{Renormalization Group functions}
\label{sec-renormalization-group-functios}

The renormalization group equation reads
\begin{equation}
\bigg(\sum_{a}\beta_{a}\frac{\partial}{\mathrm{\partial}a}-N_{F}\gamma_{\psi}-N_{A_{\mu}}\gamma_{A_{\mu}}\bigg)\Gamma^{(N_{F},N_{A_{\mu}})}=0,
\end{equation}
where $\beta_a=\Lambda(\partial a/\partial\Lambda)$, with $a=\{\Lambda, e, v_F, c, m\}$, 
are the beta functions of the parameters $e$, $v_F$, $c$, and $m$. $N_F$ and $N_{A_{\mu}}$ mean the external lines of fermions and $A_{\mu}$ field, respectively. The terms $\gamma_{\psi}$ and $\gamma_{A_{\mu}}$ are the anomalous dimension of the fermion and gauge field, respectively. However, since the polarization tensor for the gauge field is finite at one-loop (using the dimensional regularization), we can conclude that $\gamma_{A_{\mu}}=0$, therefore $\beta_{e}=0$.  Thus, for our purpose,  we just need to compute  the vertex function for the fermion, and thus, the renormalization group equation for $\Gamma^{(2,0)}$ becomes
\begin{equation}
	\left(\Lambda\frac{\partial}{\partial\Lambda}+\beta_{v_{F}}\frac{\partial}{\mathrm{\partial}v_{F}}+\beta_{m}\frac{\partial}{\mathrm{\partial}m}-2\gamma_{\psi}\right)\Gamma^{(2,0)}=0,\label{eq:RG equation}
\end{equation}
On the other hand, the vertex function $\Gamma^{(2,0)}$  can be written as
\begin{equation}
	\Gamma^{(2,0)}=\gamma_{0}p_{0}+v_{F}\gamma_{i}p_{i}-m+\Sigma(p).\label{eq:vertex function}
\end{equation}
Substituting Eq. (\ref{eq:vertex function}) into (\ref{eq:RG equation}) and grouping the terms order by order in the $1/N$ expansion, with $\beta_a=N^0 \beta_{a}^{(0)}+(1/N)\beta_a^{(1)}+\cdots$ and $\gamma_{\psi}=N^0 \gamma_{\psi}^{(0)}+(1/N) \gamma_{\psi}^{(1)}+\cdots$, we obtain
\begin{equation}
\beta_{v_{F}}=\frac{v_{F}\varUpsilon}{N}\int_{0}^{\pi}\mathrm{d}\theta\frac{2m^{2}+\Lambda^{2}v_{F}^{2}\sin^{2}\theta}{(\Lambda^{2}v_{F}^{2}+m^{2})^{2}}f(\theta,\Lambda v_{F}),
\end{equation}
\begin{equation}
\beta_{m}=\frac{2m\varUpsilon}{N}\int_{0}^{\pi}\mathrm{d}\theta\frac{\Lambda^{2}v_{F}^{2}\sin^{2}\theta+m^{2}}{(\Lambda^{2}v_{F}^{2}+m^{2})^{2}}f(\theta,\Lambda v_{F}),\label{eq:beta-m}
\end{equation}
\begin{equation}
\gamma_{\psi}=\frac{\varUpsilon}{N}\int_{0}^{\pi}\mathrm{d}\theta\frac{\Lambda^{2}v_{F}^{2}(2\sin^{2}\theta-1)+m^{2}}{(\Lambda^{2}v_{F}^{2}+m^{2})^{2}}f(\theta,\Lambda v_{F}).
\end{equation}
where we defined $\varUpsilon=-2\bar{\lambda}(\Lambda v_{F}/\pi)^{2}$.

In the next subsection we will focus only on the analysis of the band gap renormalization.

\subsection{Running mass}

To calculate the renormalized mass, we use the small-mass limit ($m^{2}\ll \Lambda^{2}v_F^2$)
to neglected the term $4m^{2}/(\Lambda^{2}v_F^2)$ in Eq. (\ref{eq:beta-m}), and
using the definition $\beta_{m}$, we get 
\begin{equation}
	m=m_{0}\exp\left[\int_{\Lambda_{0}}^{\Lambda}\frac{\mathrm{d}\Lambda}{\Lambda}C_{\bar{\lambda}}(\Lambda)\right],
	\label{eq:renormalized mass}
\end{equation}
where 
\begin{eqnarray}
C_{\bar{\lambda}}(\Lambda) & = & -\frac{4\bar{\lambda}}{\pi{}^{2}N}\int_{0}^{\pi}\mathrm{d}\theta\sin^{2}\theta\nonumber \\
 &  & \times\left[\frac{1}{1-\kappa_{21}\exp(-2z_{0}\Lambda\sin\theta)}+\bar{\lambda}\sin\theta\right]^{-1}.\qquad
 \label{eq:C}
\end{eqnarray}
For a two-dimensional system, we can show that the carrier concentration $n = N_e/A$, where $N_e$ is the number of electrons and $A$ is the area occupied by each state, can be
written as $n = N_e/A = p^2_F/\pi$ \cite{Marino-2017-livro,Salinas-2013-Livro}. Therefore, it follows that the Fermi momentum reads $p_F = (\pi n)^{1/2}$. This
plays the role of our energy scale $\Lambda$, thus, we shall use the
transformation $\Lambda /\Lambda_0\rightarrow (n/n_0 )^{1/2}$ for our renormalized functions.

We remark that, for $\epsilon_2 =\epsilon_1$ (situation illustrated in Fig. \ref{fig:luis}), Eqs. \eqref{eq:renormalized mass} and \eqref{eq:C} recover Eqs. \eqref{eq:Luis} and \eqref{eq:Luis-C} found in the literature \cite{Fernandez-etal-PRD-2020}.
For $z_0=0$ (situation illustrated in Fig. \ref{fig:luis-ext}),  Eqs. \eqref{eq:renormalized mass} and \eqref{eq:C} give Eqs. \eqref{eq:Luis} and \eqref{eq:Luis-C},  with $\epsilon_1\to (\epsilon_1+\epsilon_2)/2$.
%
\section{Application}
\label{sec-application}

Here, we apply our formulas to investigate the renormalization of the band gap for a monolayer of WSe$_{2}$, when this 2D system is inside a boron nitride substrate ($\epsilon_{1}=4$), at a distance $z_0$ from an interface with the vacuum ($\epsilon_2=1$) (see Figs. \ref{fig:nos} and \ref{modelo}).
We also investigate the opposite situation when the monolayer of WSe$_{2}$ is inside the vacuum ($\epsilon_{1}=1$), at a distance $z_0$ from the interface with the boron nitride substrate ($\epsilon_2=4$).
Taking Refs. \cite{Fernandez-etal-PRD-2020,Nguyen2019} as basis, we consider the reference values $n_{0}=1.58\times10^{12}\text{c}\text{m}^{-2}$, $\bar{\lambda}=0.48$ (when $\epsilon_1=4$), and $\bar{\lambda}=1.92$ (when $\epsilon_1=1$). 
Using Eq. \eqref{eq:renormalized mass} and putting the flavor number $N=2$ due to the degeneracy of the  valleys $K$ and $K^{\prime}$, we obtain the behavior of $m/m_0$ as function of $n$, for several situations discussed next, always comparing our results with those obtained if  Eq. \eqref{eq:Luis} (found in Ref. \cite{Fernandez-etal-PRD-2020}) were applied as if the 2D system were in a single medium $\epsilon_1$.

\begin{figure}
\centering
\subfigure[$z_0=100\text{nm}$.]{\label{Fig-4-1-100}\includegraphics[width=1.0\linewidth]{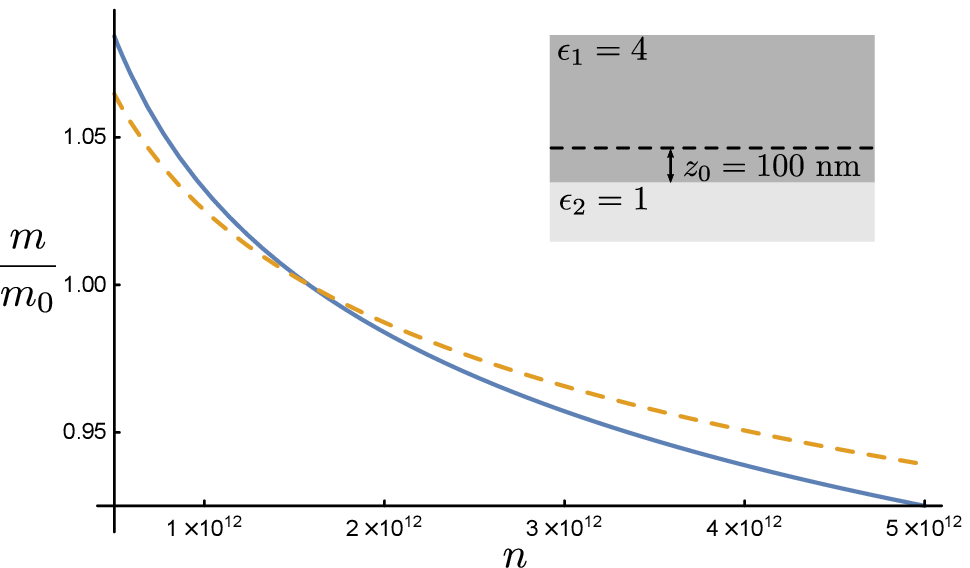}}\\
\subfigure[$z_0=300\text{nm}$.]{\label{Fig-4-1-300}\includegraphics[width=1.0\linewidth]{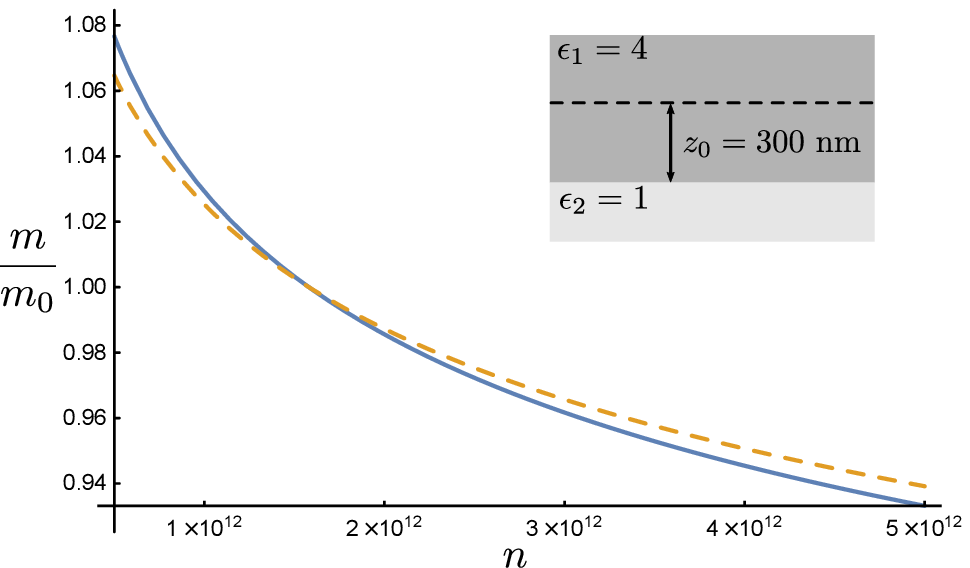}}\\
\subfigure[$z_0=500\text{nm}$.]{\label{Fig-4-1-500}\includegraphics[width=1.0\linewidth]{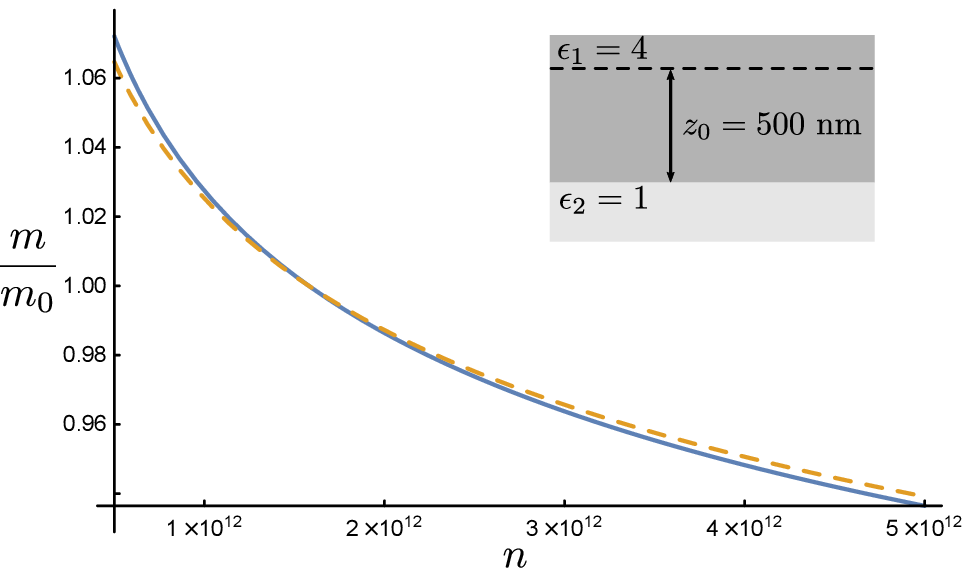}}
\caption{Ratio $m/m_0$ as function of $n$, for a monolayer of WSe$_{2}$ inside a boron nitride substrate ($\epsilon_{1}=4$) near the interface with the vacuum ($\epsilon_{2}=1$), for different values of $z_0$.  The solid (blue) and dashed (orange) lines show the curves obtained via Eqs. \eqref{eq:renormalized mass} and \eqref{eq:Luis}, respectively.}
\label{Fig-4-1}
\end{figure}

In Fig. \ref{Fig-4-1-100}, we show the ratio $m/m_0$ versus $n$, for a monolayer of WSe$_{2}$ inside a  boron nitride substrate ($\epsilon_{1}=4$), at a distance $z_0=100\text{nm}$ from an interface with the vacuum. The solid (blue) line shows the curve obtained via Eq. \eqref{eq:renormalized mass}, whereas the dashed (orange) line shows the curve obtained via Eq. \eqref{eq:Luis}, this latter applied as if the WSe$_{2}$ were immersed in an infinite boron nitride substrate, ignoring the interface with the vacuum. 
It is evident that Eq. \eqref{eq:Luis} overestimates the values of $m/m_0$ for $n>n_0$ in comparison to the precise values obtained from Eq. \eqref{eq:renormalized mass}, whereas it predicts lower values for $n<n_0$. 
On the other hand, this difference decreases as $z_0$ increases, as expected and shown in Fig. \ref{Fig-4-1-300} (for $z_0=300\text{nm}$) and Fig. \ref{Fig-4-1-500} (for $z_0=500\text{nm}$).

\begin{figure}[h]
\centering
\subfigure[$z_0=100\text{nm}$.]{\label{Fig-1-4-100}\includegraphics[width=1.0\linewidth]{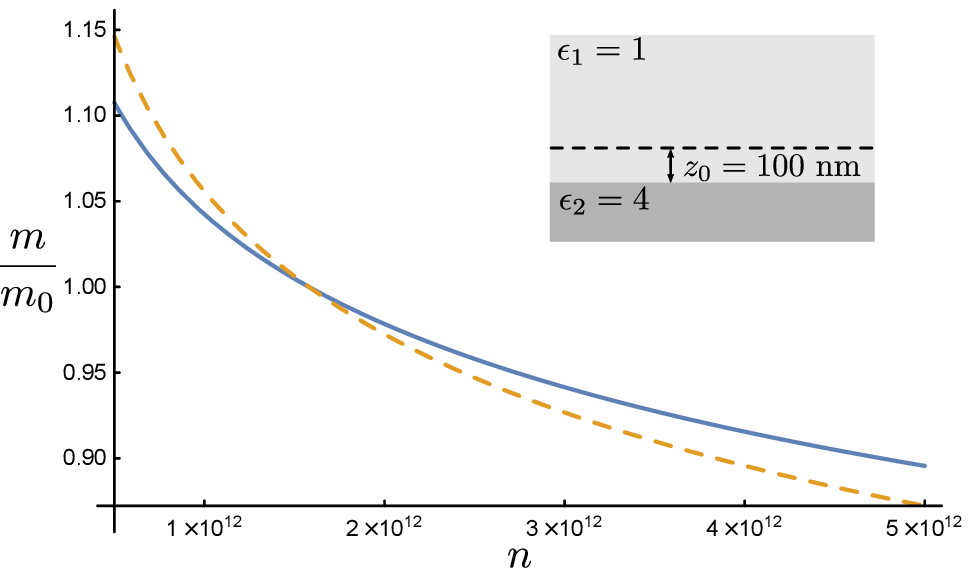}}\\
\subfigure[$z_0=300\text{nm}$.]{\label{Fig-1-4-300}\includegraphics[width=1.0\linewidth]{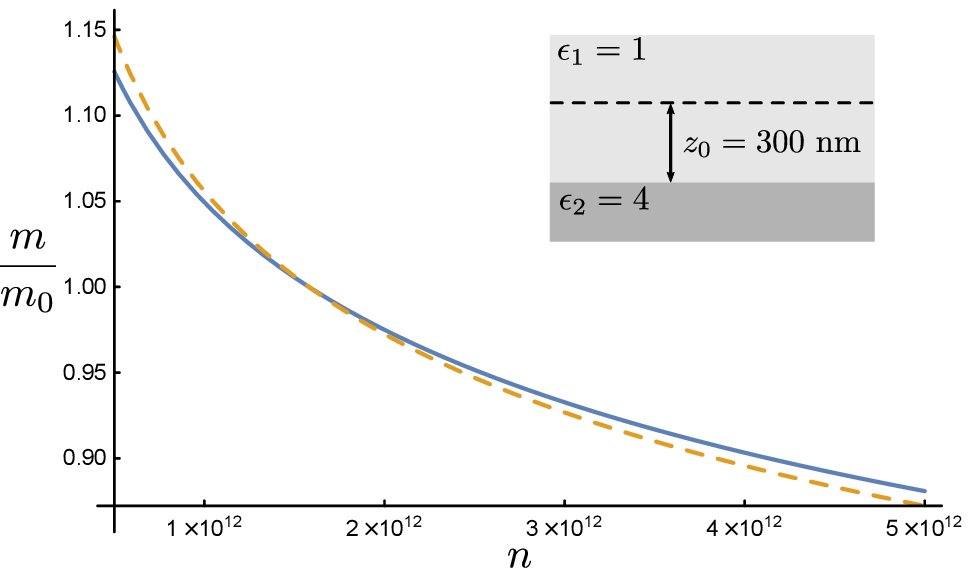}}\\
\subfigure[$z_0=500\text{nm}$.]{\label{Fig-1-4-500}\includegraphics[width=1.0\linewidth]{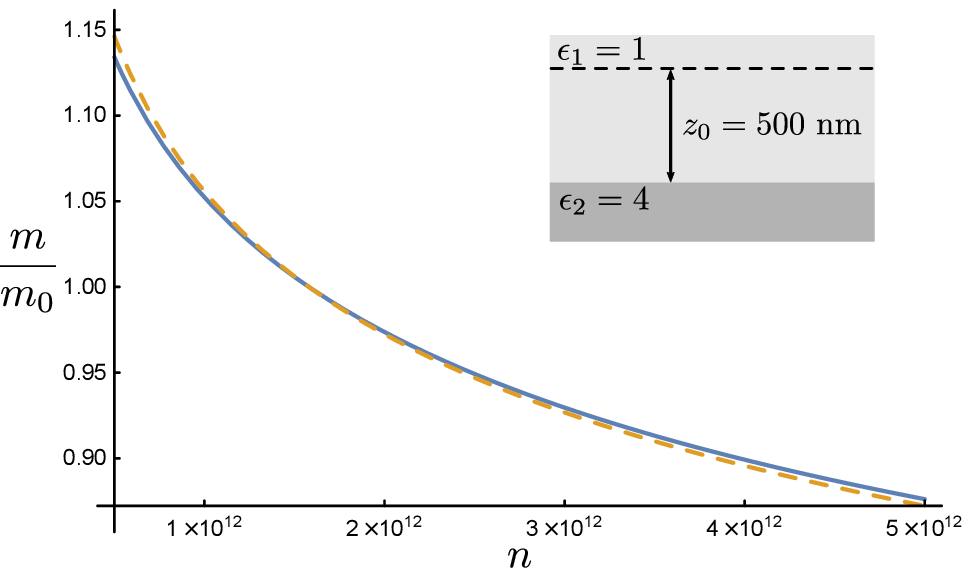}}
\caption{Ratio $m/m_0$ as function of $n$ in the opposite situation to that shown in Fig. \ref{Fig-4-1}. Here the monolayer of WSe$_{2}$ is inside vacuum ($\epsilon_{1}=1$) and near to the interface with a boron nitride substrate ($\epsilon_{2}=4$). The solid (blue) and dashed (orange) lines show the curves obtained via Eqs. \eqref{eq:renormalized mass} and \eqref{eq:Luis}, respectively.}
\label{Fig-1-4}
\end{figure}

In Fig. \ref{Fig-1-4}, we show the situation where the media are interchanged, i.e. the monolayer of WSe$_{2}$ is inside vacuum ($\epsilon_{1}=1$), at a distance $z_0$ from the interface with the  boron nitride substrate ($\epsilon_{2}=4$).
In Fig. \ref{Fig-1-4-100}, we show the ratio $m/m_0$ versus $n$ for $z_0=100\text{nm}$. 
The solid (blue) line shows the curve obtained via Eq. \eqref{eq:renormalized mass}, whereas the dashed (orange) line shows the curve obtained via Eq. \eqref{eq:Luis}, this latter applied as if the WSe$_{2}$ were immersed in an infinite vacuum, ignoring the interface with the boron nitride substrate. 
One can see that Eq. \eqref{eq:Luis} predicts lower values for $m/m_0$ in comparison to the precise values obtained from Eq. \eqref{eq:renormalized mass}, whereas it predicts higher values for $n<n_0$.
On the other hand, this difference decreases as $z_0$ increases, as expected and shown in Fig. \ref{Fig-1-4-300} (for $z_0=300\text{nm}$) and Fig. \ref{Fig-1-4-500} (for $z_0=500\text{nm}$).

\section{Final Remarks}
\label{sec-final-remarks}

We investigated, in the context of the cavity PQED and considering the RPA, the effect of a planar interface (between two nondispersive semi-infinite dielectrics) on the renormalization of the mass (band gap) in a planar 2D material located at a distance $z_0$ from the interface (Fig. \ref{fig:nos}). 
Our result [Eq. (\ref{eq:renormalized mass})] enables us to predict the behavior of the band gap as a function of the density of states, under the influence of such interface. 

As an application, we considered Eq. \eqref{eq:renormalized mass} to investigate the renormalization of the band gap for a monolayer of WSe$_{2}$, when this 2D system is inside a boron nitride substrate, at a distance $z_0$ from an interface with the vacuum (see Fig. \ref{Fig-4-1}). 
We also investigated the opposite situation when the WSe$_{2}$ is inside vacuum, at a distance $z_0$ from the interface with the boron nitride substrate (see Fig. \ref{Fig-1-4}).
We compared our results with those obtained if Eq. \eqref{eq:Luis} (found in Ref. \cite{Fernandez-etal-PRD-2020}) were applied as if the 2D system were in a single medium.
The differences between the more precise predictions, given by our Eq. \eqref{eq:renormalized mass},
and those obtained by Eq. \eqref{eq:Luis} are larger for smaller values of $z_0$ [see Figs. \ref{Fig-4-1-100} and \ref{Fig-1-4-100}], which means that the effect of the interface, captured by Eq. \eqref{eq:renormalized mass}, is significant. 
On the other hand, as $z_0$ increases, the differences decrease [see Figs. \ref{Fig-4-1-500} and  \ref{Fig-1-4-500}], which means that the effect of the interface on the renormalized band gap becomes small for larger distances $z_0$, as expected.

Since experiments verifying the dependence of the band gap with the density of states have been made \cite{Nguyen2019,Liu2019}, an experimental verification of the effects predicted here, according to Eq. \eqref{eq:renormalized mass}, seems feasible.

\begin{acknowledgments}
The authors would like to express their gratitude to Leandro O. Nascimento and Luis Fernández for their invaluable contributions and insightful discussions.
\end{acknowledgments}

\appendix
\section{Photon propagator}
\label{sec:modified-photon-propagator}

\begin{figure}[H]
	\centering
	\epsfig{file=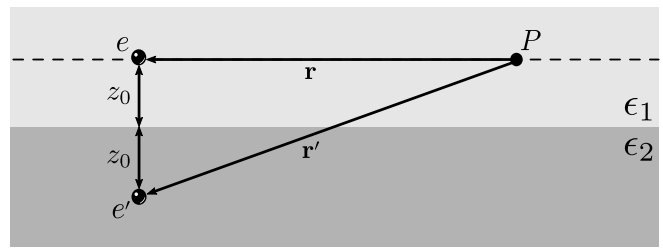,  width=1.0\linewidth}
	\caption{An electric charge $e$ and its image $e^{\prime}$, each one separated by a distance $z_0$ from the interface between the two dielectric media. The charge $e$ is in the medium with dielectric constant $\epsilon_{1}$ and its image is in the medium with dielectric constant $\epsilon_{2}$. $P$ is an arbitrary point of the plane that contains the charge $e$ and it is parallel to the interface.}
	\label{modelo_JD}
\end{figure}

Consider an electric charge $e$ in a semi-infinity
dielectric medium with dielectric constant $\epsilon_{1}$ separated by a
distance $z_{0}$ from the interface between another semi-infinity dielectric
medium with dielectric constant $\epsilon_{2}$.  By the image method, the
potential $V$ for this configuration at an arbitrary point $P$ (also separated by a
distance $z_{0}$ from the interface) is (see Fig. \ref{modelo_JD})
\begin{equation}
V(|\mathbf{r}|)=\frac{1}{4\pi\epsilon_{1}}\left(\frac{e}{|\mathbf{r}|}+\frac{e^{\prime}}{|\mathbf{r}^{\prime}|}\right),
\label{eq:VP}
\end{equation}
where $|\mathbf{r}|$ is the distance between the charge $e$ and
the point $P$ and $|\mathbf{r}^{\prime}|$ is the distance between
the image charge $e^{\prime}$ and the point $P$. The
image charge $e^{\prime}$ given by
\begin{equation}
e^{\prime}=-\kappa_{21}e,
\end{equation}
and the distance between the image charge and the point $P$ is
$|\mathbf{r^{\prime}}|=\sqrt{|\mathbf{r}|^{2}+4z_{0}^{2}}$.
Therefore, the potential at $P$ is
\begin{equation}
V(|\mathbf{r}|)=\frac{e}{4\pi\epsilon_{1}}\left(\frac{1}{|\mathbf{r}|}-\frac{\kappa_{21}}{\sqrt{|\mathbf{r}|^{2}+4z_{0}^{2}}}\right).
\end{equation}
From this static potential, via the inverse Fourier transform, which is given by (see, for instance, Ref. \citep{Roberts-Williams-ProgPartNuclPhys-1994})
\begin{equation}
\Delta_{00}^{(0)}(|\mathbf{k}|)=\frac{1}{e^{2}}\int\mathrm{d}^{2}\mathbf{r}\,\mathrm{e}^{-i\mathbf{k}\cdot\mathbf{r}}eV(|\mathbf{r}|),
\end{equation}
where $k_{0}=0$, and $\mathbf{k}$
and $\mathbf{r}$ are restricted to the plane $z=z_{0}$, thus
we can write
$\mathbf{k}\cdot\mathbf{r}=|\mathbf{k}||\mathbf{r}|\cos\varphi$ and $\mathrm{d}^{2}\mathbf{r}=|\mathbf{r}|\,\mathrm{d}\varphi\,\mathrm{d}|\mathbf{r}|$.
After performing the integration, we obtain Eq. (\ref{eq:gauge-prop-dieletrico}).
The first term is the propagator in the absence of the medium with
$\epsilon_{2}$ (recovered when $z_{0}\rightarrow\infty$), and
the exponential term arises due to the presence of the medium with $\epsilon_{2}$.

%

\end{document}